\documentclass[iop]{emulateapj}

\usepackage{times,natbib,graphicx,amsmath}

\shorttitle{The Hard State of 1RXS J1804}
\shortauthors{Ludlam et al.}

\begin{document}

\title{$\emph{NuSTAR}$  and $\emph{XMM-Newton}$ Observations of the Neutron Star X-ray Binary 1RXS J180408.9-34205}
\author{R. M. Ludlam\altaffilmark{1},
J. M. Miller\altaffilmark{1}, E. M. Cackett\altaffilmark{2}, A. C. Fabian\altaffilmark{3}, M. Bachetti\altaffilmark{4}, M. L. Parker\altaffilmark{3}, J. A. Tomsick\altaffilmark{5}, D. Barret\altaffilmark{6,7}, L. Natalucci\altaffilmark{8}, V. Rana\altaffilmark{9}, F. A. Harrison\altaffilmark{9}}
\altaffiltext{1}{Department of Astronomy, University of Michigan, 1085 South University Ave, Ann Arbor, MI 48109-1107, USA}
\altaffiltext{2}{Department of Physics \& Astronomy, Wayne State University, 666 W. Hancock St., Detroit, MI 48201, USA}
\altaffiltext{3}{Institute of Astronomy, Madingley Road, Cambridge CB3 0HA}
\altaffiltext{4}{INAF/Osservatorio Astronomico di Cagliari, via della Scienza 5, I-09047 Selargius (CA), Italy}
\altaffiltext{5}{Space Sciences Laboratory, 7 Gauss Way, University of California, Berkeley, CA 94720-7450, USA}
\altaffiltext{6}{Universit de Toulouse; UPS-OMP; IRAP; Toulouse, France}
\altaffiltext{7}{CNRS; Institut de Recherche en Astrophysique et Plantologie; 9 Av. colonel Roche, BP 44346, F-31028 Toulouse cedex 4, France}
\altaffiltext{8}{Istituto Nazionale di Astrofisica, INAF-IAPS, via del Fosso del Cavaliere, 00133 Roma, Italy}
\altaffiltext{9}{Cahill Center for Astronomy and Astrophysics, California Institute of Technology, Pasadena, CA 91125}
\begin{abstract} 
We report on observations of the neutron star (NS) residing in the low-mass X-ray binary 1RXS J180408.9-34205 taken 2015 March by $\emph{NuSTAR}$ and $\emph{XMM-Newton}$ while the source was in the hard spectral state. We find multiple reflection features (Fe K$_{\alpha}$ detected with $\emph{NuSTAR}$; N VII, O VII, and O VIII detected in the RGS) from different ionization zones. Through joint fits using the self consistent relativistic reflection model {\sc relxill}, we determine the inner radius to be $\leq 11.1\ R_{g}$. For a 1.4 M$_{\odot}$ NS with a spin of $a_{*}=0$, this is an inner disk radius of $\leq22.2$ km. We find the inclination of the system to be between $18^{\circ}$-$29^{\circ}$. If the disk is truncated at a radius greater than the neutron star radius, it could be truncated by a boundary layer on the neutron star surface. It is also possible that the disk is truncated at the magnetospheric radius; conservative estimates would then imply $B\leq(0.3 -1.0)\times10^{9}$ G at the magnetic poles, though coherent pulsations have not been detected and the source is not identified as a pulsar.
\end{abstract}

\section{Introduction}
Low-mass X-ray binaries (LMXBs) consist of a compact object that accretes matter via Roche-lobe overflow from a roughly stellar mass companion. Broad iron line profiles have been seen in these accreting systems that harbor a black hole (BH; e.g. \citealt{Fabian89}; \citealt{miller07}; \citealt{reis08}, \citeyear{reis09a}) or a neutron star  (NS; e.g. \citealt{BS07}; \citealt{papitto08}; \citealt{cackett08}, \citeyear{cackett09}, \citeyear{cackett10}; \citealt{disalvo09}; \citealt{Egron13};\citealt{miller13}) as the primary accreting compact object.  The asymmetrically, broadened profile of the Fe K$_{\alpha}$ line gives a direct measure of the position of inner disk since the effects of gravitational redshift and Doppler redshift/boosting on the emission line become stronger closer to the compact object \citep{Fabian89}. 

The Fe K$_{\alpha}$ line in NS LMXBs can be used to set an upper limit for the radius of NS since the disk must truncate at or before the surface (\citealt{cackett08}, \citeyear{cackett10}; \citealt{reis09b}; \citealt{miller13}; \citealt{degenaar15}). Constraining the radius of the NS can, in turn, lead to narrowing down the equation of state of the cold, dense matter under extremely dense physical conditions \citep{lattimer11}. 
 On the other hand, if the accretion rate is low, it may be possible to constrain the stellar magnetic field based on the radius at which the disk truncates (\citealt{cackett09}; \citealt{Pap09}; \citealt{miller11}; \citealt{degenaar14}).

1RXS J180408.9-34205 (hereafter 1RXS J1804) was classified as a neutron star LMXB after Type I X-ray bursts were detected during the INTEGRAL Galactic Bulge Monitoring in 2012 May \citep{ATel.6997}. The bursts provided an upper limit of 5.8 kpc as the distance to the binary system. Follow up observations with $\emph{Swift}$ less than a month later showed that the neutron star had returned to quiescence \citep{ATel.4085}. 

The source remained quiet until early 2015 when it began to show a steady increase of hard X-rays in the 15-50 keV as seen with $\emph{Swift}$/BAT \citep{ATel.6997}. MAXI/GSC also detected a gradual increase in the 2-20 keV band \citep{ATel.7008} confirming that 1RXS J180408 had indeed gone into outburst. The hard X-ray brightness increased until plateauing  at 100 mCrab on 2015 February 5 with a flux $\sim160$ times the quiescent rate \citep{ATel.7039}. Type I X-ray bursts were again seen during this observation \citep{ATel.7039}, reaffirming the source's classification as having a neutron star as its primary. \citet{ATel.7039} noted that the X-ray spectrum seen was adequately described by a simple absorbed power law model. 

Radio observations performed with the VLA in 2015 March detected the presence of a jet \citep{ATel.7255}. $\emph{Swift}$ X-ray spectra were still described by a single power law placing J180408's location on the radio/X-ray luminosity plane consistent with hard-state neutron star LMXBs \citep{ATel.7255}. However, in early April a drop in the hard X-rays was seen with $\emph{Swift}$/BAT in conjugation with an increase in softer X-ray emission seen by MAXI/GSC \citep{ATel.7352}. This suggests that the source entered a soft spectral state. $\emph{Swift}$ spectra could no longer be described by a simple power law but required the presence of an additional disk component and black body component \citep{ATel.7352}.  

We report on observations taken 2015 March by $\emph{NuSTAR}$ and $\emph{XMM-Newton}$ while the source was still in the hard spectral state. We focus on the importance of constraining the neutron star's inner radius from its reflection spectrum in this state since it is not obscured by thermal emission in the lower energies.

\section{Observations and Data Reduction}
\subsection{$\emph{NuSTAR}$}
$\emph{NuSTAR}$ \citep{nustar} observed the target for $\simeq$80 ks between 9:21 UT on 2015 March 5 and 7:36 UT March 6 (Obs ID 80001040002). Lightcurves and spectra were created using 120$^{\prime \prime}$ circular extraction region centered around the source and another region of the same dimensions away from the source as a background using the {\sc{nuproducts}} tool from {\sc nustardas} v1.3.1 with {\sc caldb} 20150123. There were a total of 5 Type 1 X-ray bursts present in the lightcurves. We will report on the bursts in a separate paper. We created good time intervals (GTIs) to remove 100 -- 170 s after the initial fast rise (depending on the duration of the individual burst) to separate these from the steady emission. These GTIs were applied during the generation of the spectra for both the FPMA and FPMB. Initial modeling of the steady spectra with a constant fixed to 1 for the FPMA and allowed to float for the FPMB was found to be within 0.95-1.05. We proceeded to combine the two source spectra, background spectra, and ancillary response matrices via {\sc addascaspec}. We use {\sc addrmf} to create a single redistribution matrix file. The spectra were grouped using {\sc grppha} to have a minimum of 25 counts per bin \citep{cash}. The net count rate was 39.9 counts/s.

\subsection{$\emph{XMM-Newton}$}
$\emph{XMM-Newton}$ observed the target on 2015 March 6 (Obs ID 0741620101) on revolution 2791 for 57 ks. The EPIC-pn camera was operated in \lq timing' ($\sim$41.1 ks) and \lq burst' ($\sim$9.5 ks) mode during this time with a medium optical blocking filter. The RGS cameras were also operational for $\simeq$55.8 ks. The data were reduced using SAS ver. 13.5. We are primarily interested in the spectra obtained from RGS due to the high-resolution. Moreover, the Epic-pn \lq timing' mode and \lq burst' mode spectra suffer from pile-up and calibration issues \citep{walton12}. The net count rate was 210 counts/s and 167 counts/s for the \lq timing' and \lq burst' mode, respectively. Even when eliminating the central most columns from the observation, the spectra obtained in the different modes did not agree with $\emph{NuSTAR}$ or even each other. $\emph{NuSTAR}$ does not suffer from photon pile-up, so we chose to characterize the spectrum through the Fe K band using only $\emph{NuSTAR}$.
During the observation, there were 7 Type 1 X-ray bursts which were filtered from the data by GTIs that removed 125 -- 225 s after the initial fast rise depending on the duration of each burst. We then created spectra through {\sc rgsproc} that were grouped using {\sc grppha} to have a minimum of 25 counts per bin to allow the use of $\chi^{2}$ statistics. There were an average of 2.6 counts/s for RGS1 and 3.3 counts/s for RGS2.

\section{Spectral Analysis and Results}
We use XSPEC version 12.8.1 \citep{arnaud96} in this work. All errors are quoted at $\geq$ 90$\%$ confidence level. 
We account for the neutral column hydrogen density along the line of sight via tbnew\footnote{Wilms, Juett, Schulz, Nowak, in prep, http://pulsar.sternwarte.uni-erlangen.de/wilms/research/tbabs/index.html}. The solar abundance was set to {\sc wilm} \citep{wilms00} and {\sc vern} cross sections \citep{Vern96} were used. The RGS and $\emph{NuSTAR}$ were considered between 0.45-2.1 keV and 3.5-50.0 keV, respectively. The choice of the lower energy bound of $\emph{NuSTAR}$ data was motivated by two high bins in the low energy of the spectrum. 

\begin{figure}
\centering
\includegraphics[angle=270,width=8.4cm]{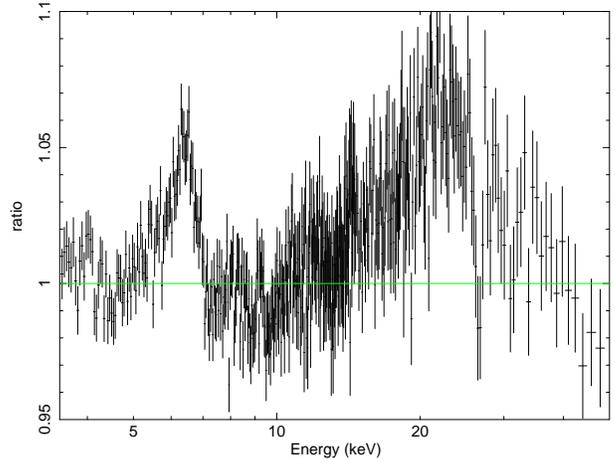}
\caption{Ratio of the data to the continuum model for $\emph{NuSTAR}$ observation of 1RXS J180408. A simple cut-off power law was fit over the energies of 3.5-5 keV, 8-15 keV, and 30.0-50.0 keV. The iron line region from 5-8 keV and reflection hump was ignored to prevent the feature from skewing the fit. We ignore the region above 15 keV in this fit due to the presence of a X-ray reflection hump that peaks between 20-30 keV. The data was rebinned for plotting purposes. 
}
\label{fig:feline}
\end{figure}

\subsection{NuSTAR}
We start with investigating the $\emph{NuSTAR}$ data alone. Initial fits were performed with an absorbed power law with two Gaussian lines at 10.1 keV and 11.5 keV to account for instrumental response features \citep{nustar}. These Gaussians will be present in all $\emph{NuSTAR}$ fits hereafter. This gives a particularly poor fit ($\chi^{2}/dof = 9545/1159$). The addition of a blackbody to account for thermal component in the spectrum did not improve the goodness of fit ($\chi^{2}/dof = 9545/1157$). The absence of a thermal component is consistent previous investigations of 1RXS J1804 in the hard state (\citealt{ATel.7008}; \citealt{ATel.7039}; \citealt{ATel.7255}). 

A prominent Fe K$_{\alpha}$ line centered $\sim 6.4$ keV with a red wing extending to lower energies and Fe edges can be seen rising above the continuum in Figure 1. There is also a prominent reflection hump in the higher energies above 10 keV.  These reflection features not being properly described lend to the poor goodness of fit. Since a thermal component is not necessary to describe the spectrum, the only source of emission is a power law. This means the disk is being illuminated by a source of hard X-ray photons and likely follows a radial dependence $R^{-3}$ profile.

To properly describe the reflection features and relativistic effects present in the data we employ the model {\sc relxill} \citep{garcia14}. This model self-consistently models for X-ray reflection and relativistic ray tracing for a power law irradiating an accretion disk. The parameters of this model are as follows: inner emissivity ($q_{in}$), outer emissivity ($q_{out}$), break radius ($R_{break}$) between the two emissivities, spin parameter ($a_{*}$), inclination of the disk ($i$), inner radius of the disk ($R_{in}$), outer radius of the disk ($R_{out}$), redshift ($z$), photon index of the power law($\Gamma$), log of the ionization parameter ($log(\xi)$), iron abundance ($A_{Fe}$), cut off energy of the power law ($E_{cut}$), reflection fraction ($f_{refl}$), angleon, and normalization.  

A few reasonable conditions were enforced when making fits with {\sc relxill}. First, we tie the outer emissivity index, $q_{out}$ to the inner emissivity index, $q_{in}$, to create a constant emissivity index that is allowed to vary between 1 to 3. Next, we fix the spin parameter, $a_{*}$ (where $a_{*}=cJ/GM^{2}$), in the model {\sc relxill} to 0 in the subsequent fits since NS in LMXBs have $a_{*} \leq 0.3$ (\citealt{miller11}; \citealt{Galloway08}). This does not hinder our estimate of the inner radius since the position of the inner most stable circular orbit (ISCO) is relatively constant for low spin parameters. In fact, \citet{Miller98} found that corrections for frame-dragging for $a_{*}<0.3$ give errors $\ll10\%$. Further, the outer disk radius has been fixed to 400 $R_{g}$ (where  $R_{g} = GM/c^{2}$). Lastly, to ensure that the inclination is properly taken into account in the reflection from the disk, we set angleon=1 rather than 0 (angle averaged). 

Fits with {\sc tbnew$*$relxill} and two Gaussian components, again account for instrument response features, provide a significantly better fit ($\chi^{2}/dof=1213/1151$). This is well over a 20$\sigma$ improvement. Values for model parameters are given in Table 1. Figure 2 shows the modeled spectrum and ratio of the model to the data. The absorption column density is $(2.0\pm2.0)\times 10^{21}$ cm$^{-2}$. The large uncertainties are due to the lack of data in the low energy range where X-ray absorption is prevalent. The photon index is consistent with the hard limit of 1.4 at the 90\% confidence level. The power law cut off energy is $45_{-1}^{+2}$ keV. The inclination of the disk is $25^{\circ}\pm4^{\circ}$. 
The parameter of most interest is the inner radius, $R_{in}$, which was found to be $1.9_{-0.8}^{+1.4}$ ISCO.

\begin{table*}
\caption{Relxill Fitting of $\emph{NuSTAR}$ and RGS 
}
\label{tab:relxill} 
\begin{minipage}{160mm}
\begin{center}
\begin{tabular}{llccc}
\hline
Component & Parameter & NuSTAR & RGS & NuSTAR $+$ RGS \\
\hline
{\sc tbnew}
&$N_\mathit{H} (10^{22})$
&$0.2\pm0.2$
&$0.360\pm0.002$
&$0.345\pm0.001$
\\
& $A_{O}$
&...
&$1.77\pm0.01$
&$1.71\pm0.01$
\\
{\sc relxill}
&$q_{in}=q_{out}$
&$2.4_{-0.4}^{+0.5}$
&$<1.9$
&$1.3\pm0.2$
\\
&$a_{*} '$
&0
&0
&0
\\
&$\mathit{i} (^{\circ})$
&$25\pm4$
&$18.4_{-0.2}^{+0.5}$
&$18.3\pm0.2$
\\
&$R_\mathit{in} (ISCO) $
&$1.9_{-0.8}^{+1.4}$
&$1.2_{-0.2}^{+3.3}$
&$\leq1.85$
\\
&$R_\mathit{out} (R_\mathit{g}) '$ 
&400
&400
&400
\\
&$\mathit{z} '$
&0
&0
&0
\\
&$\Gamma$
&$<1.424$
&$<1.402$
&$1.402\pm0.001$
\\
&$log(\xi)$
&$2.4_{-0.1}^{+0.4}$
&$3.14\pm0.03$
&$2.75\pm0.01$
\\
&$A_\mathit{Fe}$
&$0.83_{-0.3}^{+0.4}$
&$<0.51$
&$<0.51$
\\
&$E_\mathit{cut} (keV)$
&$45_{-1}^{+2}$
&$30\pm3$
&$47.3\pm0.3$
\\
&$\mathit{f}_\mathit{refl}$
&$0.15\pm0.02$
&$0.21\pm0.01$
&$0.15\pm0.01$
\\
&angleon$'$
&1
&1
&1
\\
&norm $(10^{-2})$
&$9.5_{-0.6}^{+0.3}$
&$9.81\pm0.04$
&$9.35\pm0.01$
\\
{\sc relline}
&$E_\mathit{line}$
&...
&$0.569\pm0.001$
&$0.569\pm0.001$
\\
&norm $(10^{-3})$
&...
&$6.7\pm0.8$
&$3.9\pm0.6$
\\
\hline
&$\chi_\nu^{2}$(dof)
&1.05 (1151)
&1.48 (2804)
&1.34 (3963)
\\
\hline
$'$ = fixed
\end{tabular}

\medskip
Note.--- Errors are quoted at $\geq$ 90 \% confidence level. Setting angleon=1 takes the inclination into account when modeling reflection. A constant was allowed to float between $\emph{NuSTAR}$ and the RGS data. The $\emph{NuSTAR}$ was frozen at the value of 1.0, RGS1 was fit at $1.234\pm0.006$, and RGS2 was fit at $1.182\pm0.005$. The emissivity index, inclination, and inner radius in {\sc relline} were tied to the values in {\sc relxill}. The unabsorbed flux from the combined $\emph{NuSTAR}$+RGS fit from 0.45-50.0 keV is $F_{unabs, (0.45-50.0\ \mathrm{keV})}=1.71\times10^{-9}$ ergs cm$^{-2}$ s$^{-1}$. This gives a luminosity of $L_{(0.45-50.0\ \mathrm{keV})}=6.8\times10^{36}$ erg s$^{-1}$ ($\sim3-4$\% L$_{\mathrm{Edd}}$).

\end{center}
\end{minipage}
\end{table*}

\begin{figure}
\centering
\includegraphics[angle=270,width=8.4cm]{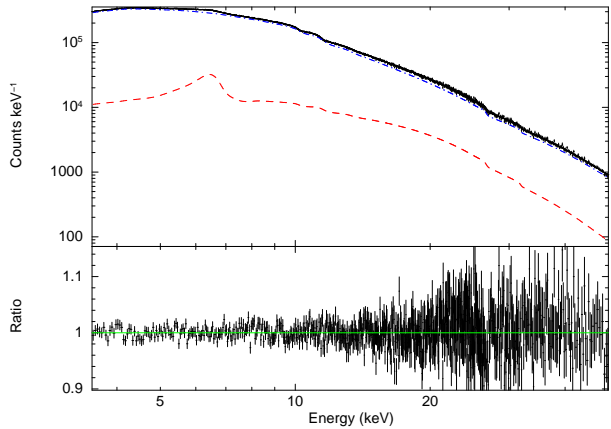}
\caption{$\emph{NuSTAR}$ spectrum fit with {\sc tbnew$*($relxill$+$gauss$+$gauss}) ($\chi_{red}^{2}=1.05$). Gaussians were fit at 10.1 keV and 11.5 keV to account for instrumental response features. The red dashed line shows the reflection spectrum from the {\sc relxill} model. The blue dot-dash line shows the power law component within the {\sc relxill} model. The lower panel shows the ratio of the data to model. Spectrum has been rebinned for plotting purposes. Parameter values can be seen in the third column of Table 1.}
\label{fig:relxill}
\end{figure}  

\subsection{RGS}
The RGS spectra showed prominent broad features rising above the continuum emission (modeled with an absorbed power law) at $\sim 0.50$ keV, $\sim 0.57$ keV, and $\sim 0.65$ keV corresponding to H-like N VII, He-like O VII, and H-like O VIII, respectively (see Figure 3). Applying simple Gaussian lines place their centroid energy at $0.501\pm0.002$ keV, $0.568\pm0.002$ keV, and $0.655\pm0.001$ keV  detected at the 3.2$\sigma$, 5.9$\sigma$, and 9$\sigma$ confidence level. To test the robustness of these lines to different assumptions about the ISM we applied {\sc ismabs} \citep{gatuzz14} with the same power law continuum. The abundance of each element is left as a free parameter so that the photoelectric absorption edges in {\sc ismabs} have the greatest possible opportunity to model the data in a way that would remove line-like features. The lines are still present in the spectra though their significance has decreased slightly with N VII line only being marginally significant. We believe the N VII to be a real feature but due to its location at the edge of the effective area of the detector we are unable to model it properly in subsequent fits. See Table 2 for the values obtained for the Gaussian lines with each absorption model. 

\begin{table*}
\caption{Emission Line in RGS Spectra}
\label{tab:comp}
\begin{minipage}{160mm}
\begin{center}
\begin{tabular}{lcccccc}
\hline
\multicolumn{1}{c}{Abs Model}
&\multicolumn{1}{c}{Ion}
&\multicolumn{1}{c}{Lab E (keV)}
&\multicolumn{1}{c}{E$_{centroid}$ (keV)}
&\multicolumn{1}{c}{$\sigma$ (keV) ($10^{-2}$)}
&\multicolumn{1}{c}{norm ($10^{-4}$)}
&\multicolumn{1}{c}{ Significance ($\sigma$)}
\\
\hline
{\sc tbnew}
&N VII & 0.5003 & $0.501\pm0.002$ &$0.6\pm0.2$& $1.6\pm0.5$&$3.2$\\
&O VII & 0.574 & $0.568\pm0.002$ & $1.1\pm0.2$& $2.9\pm0.5$& $5.9$\\
&O VIII & 0.654 & $0.655\pm0.001$& $1.2\pm0.2$& $4.5\pm0.5$& $9$\\
\hline
{\sc ismabs}
&N VII & 0.5003 & $0.501\pm0.002$ &$0.6\pm0.3$& $1.3\pm0.5$&$2.6$\\
& OVII & 0.574 & $0.571\pm0.001$ & $0.8\pm0.2$ & $2.2\pm0.5$ & $4.4$\\
& O VIII &  0.654 & $0.654\pm0.001$&$1.0\pm0.2$& $3.4\pm0.6$ & $5.7$\\
\hline
\end{tabular}

\medskip
Note.--- In {\sc tbnew}, all abundances were set to solar with $N_{H}=2.0 \times 10^{21}$ cm$^{-2}$ \citep{dl90}. Normalization is given in units of photons cm$^{-2}$ s$^{-1}$. An absorbed power law was used to model the continuum. In {\sc ismabs}, elemental abundances were allowed to vary. The continuum was modeled with a power law of photon index 1.4.
\end{center}
\end{minipage}
\end{table*}

We then applied the self consistent reflection model {\sc relxill} in combination with {\sc tbnew} to describe the lines simultaneously. We allowed the abundance of O to be a free parameter in {\sc tbnew} to fully account for the O VIII K-edge at 0.53 keV that can be seen in Figure 3. The model was not able to account for the He-like O VII line with the other two H-like lines, likely due to the lower ionization parameter needed to create the line. We employ the relativistic line model {\sc relline} to describe this feature. The emissivity index, inclination, and inner radius were tied to the parameters in {\sc relxill}. The limb parameter was set to 2 for consistency with {\sc relxill} which assumes limb brightening. We only allowed the normalization and line energy to be free. The addition of the {\sc relline} component to {\sc relxill} improves the overall fit by $\Delta \chi^{2}=147$ (for 2 d.o.f.). The resulting best fit can be seen in Table 1. The probability of the additional component improving the fit by chance was found to be negligibly small ($3.19\times10^{-24}$) via an F-test. This is a 9.6$\sigma$ improvement. The unfolded model can be seen in Figure 4. Figure 5 shows the {\sc relxill} only model decomposed into the continuum and reflection components that are comprised in the model.

The cut energy and emissivity are not well constrained likely due to limited band width. Again, the photon index is consistent with the hard limit of 1.4 at the 90\% confidence level. The absorption column is better constrained than in the previous fit with $N_{H}=3.60\pm0.02 \times 10^{21}$ cm$^{-2}$ with an oxygen abundance $1.77\pm0.01$ solar. The reflection fraction is higher than that of the $\emph{NuSTAR}$ spectrum meaning it is dominated by more reflection. The high reflection fraction reaffirms that the emission lines are reflection features. The inclination is several degrees lower than the inclination found in the higher energy range, but is generally consistent. The location of the inner disk ($1.2_{-0.2}^{+3.3}$ ISCO) is compatible with the iron line region. 

\begin{figure}
\centering
\includegraphics[angle=270,width=8.4cm]{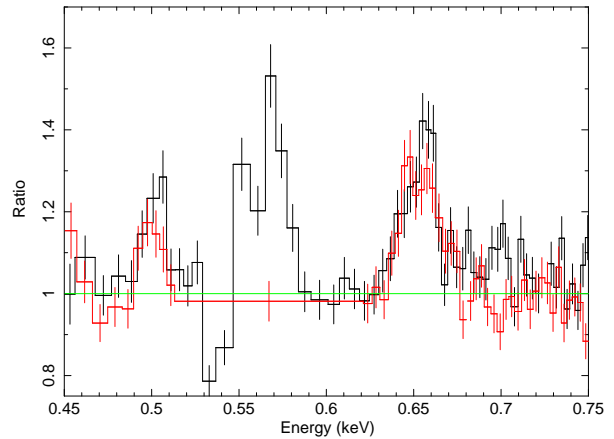}
\caption{Ratio of the data to the continuum model for $\emph{XMM-Newton}$ RGS1 (black) and RGS2 (red) showing three emission lines (N VII, O VII, O VIII) and O VIII K edge. The continuum was modeled with a simple absorbed power law was fit over the energies of 0.45-2.1 keV. The N VII line is located near the edge of the effective area of the detector and therefore not modeled in subsequent fits. The data was rebinned and the x-axis was rescaled for plotting purposes. 
}
\label{fig:rgsline}
\end{figure}

\begin{figure}
\centering
\includegraphics[angle=270,width=8.4cm]{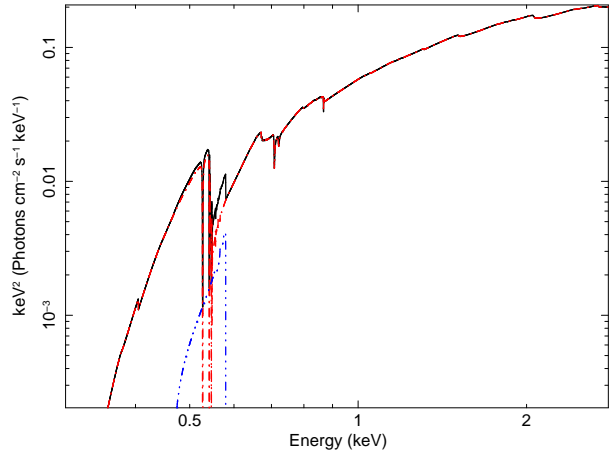}
\caption{Unfolded model spectrum for  $\emph{XMM-Newton}$ RGS. The blue component corresponds to {\sc relline}. The red component illustrates {\sc relxill} which includes the power law. 
}
\label{fig:rgsmo}
\end{figure}

\begin{figure}
\centering
\includegraphics[angle=270,width=8.4cm]{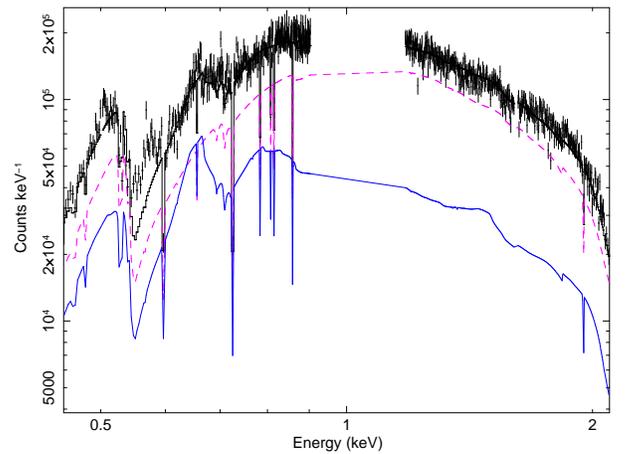}
\caption{ Decomposed {\sc relxill} model spectrum for  $\emph{XMM-Newton}$ RGS1 showing the reflection component (solid blue) and power law continuum (dashed purple). The {\sc relline} component is not shown to illustrate the models ability to fit the O VIII line.
}
\label{fig:rgsmo}
\end{figure}

\subsection{NuSTAR+RGS}
Due to the limited energy bandwidth of the RGS, we fit the $\emph{NuSTAR}$ and RGS together in order to obtain better estimates of the spectral parameters of interest. 
Tying the RGS data with the $\emph{NuSTAR}$ data provides an opportunity to fit multiple relativistic reflection features over a wide energy range ($\sim$0.45-50.0 keV) and from different ionized regions in the disk. A constant was allowed to float between the two observations, but all other model parameters were tied. The constant was frozen at the value of 1.0 for $\emph{NuSTAR}$ and free to vary for RGS1 and RGS2. The constant for RGS1 was  found to be $c=1.234\pm0.006$ and $c=1.182\pm0.005$ for RGS2.  The model we first use is {\sc tbnew$*($relxill$+$relline$+$gauss$+$gauss)} to model the relativistic reflection features with X-ray absorption along the line of sight. The Gaussian lines are to account for instrumental response features. See Table 1 and  Figure 6.

Simultaneous fits return a neutral absorption column density $N_{H}=(3.45\pm0.01)\times10^{21}$  cm$^{-2}$ with a slight over abundance in oxygen, O$=1.71\pm0.01$ solar. The emissivity index, $q=1.3\pm0.2$, suggests that a large area of the disk is being illuminated by a hard X-ray source. The ionization parameter being $log(\xi)=2.75\pm0.01$ is within reason for an accreting source in the LHS \citep{cackett10}, although there are a number of cases where lower ionizations are found for a NS in the LHS (SAX J1808.4-3658: \citealt{Pap09}; \citealt{cackett09}, 4U 1705-44: \citealt{dai10}; \citealt{disalvo15}). The iron abundance was allowed to be a free parameter and is consistent with the hard limit of 0.5 at the 90\% confidence level. The photon index of the power law component that is responsible for producing the reflection spectrum is on the harder side with $\Gamma=1.402\pm0.001$. The reflection fraction, $\mathit{f}_\mathit{refl}$, is $0.15\pm0.01$. The inclination agrees with the previous individual fit to the RGS data alone. The inner radius, $R_{in}\leq1.85$ ISCO, is consistent within uncertainties with the individual fits to $\emph{NuSTAR}$ and RGS alone. Figure 7 shows the change in $\chi^{2}$ when stepping the inner disk radius parameter out to 5 ISCO using the \lq \lq steppar" command in XSPEC.

To explore the multiple ionization parameters that seem prevalent in the RGS data, we remove the {\sc relline} component and apply a double {\sc relxill} model to the joint spectra. The idea is to be able to map the X-ray interactions with accretion disk radii. We tie the emissivity index, inclination, inner radius, photon index, and iron abundance between the two {\sc relxill} components and allow the ionization, reflection fraction, and normalization to change between them. Parameter values can be seen in Table 3.

The higher ionization parameter agrees with the values obtained in the previous fits. The lower ionization value accounts for the production of the O VII line. The inner radius, $R_{in}<4.1$ ISCO, agrees with the value obtained in the previous model when using {\sc relline} to describe the O VII line, which does not make assumptions about the underlying gas physics like {\sc relxill} does. The reflection fraction for each component are consistent with one another, though the normalization is not well constrained.
 We do not expect these two different ionization states to occur at the same radii, and as such, tried to free the inner radius in each of the {\sc relxill} components to map the radius at which the X-ray interactions with the disk occurred. We were unable to obtain radii that were statistically distinct.
The goodness of fit of the double {\sc relxill} model is comparable to the simpler fit with just one {\sc relxill} component and {\sc relline}. We chose to use the parameter values from the simpler model fits in subsequent calculations. 

If the inner disk is truncated substantially above the neutron star surface itself, as allowed by our upper limits on $R_{in}$, then the disk may instead be truncated by a boundary layer extending from the stellar surface. The upper limit on $R_{in}$ would require a boundary layer that is a few times the stellar radius in size. Alternatively, the disk could be truncated by magnetic pressure.
We can place an upper limit on the strength of the field using the upper limit of $R_{in}=1.85$ ISCO. Assuming a mass of 1.4 M$_{\odot}$, taking the distance to be 5.8 kpc \citep{ATel.6997}, and using the unabsorbed flux from 0.45-50.0 keV of $1.71\times10^{-9}$ erg cm$^{-2}$ s$^{-1}$ as the bolometric flux, we can determine the magnetic dipole moment, $\mu$, from Equation (1) taken from \citet{cackett09}. 
\begin{equation}
\begin{aligned}
\mu = 3.5 \times 10^{23} \ k_{A}^{-7/4} \ x^{7/4} \left(\frac{M}{1.4 M_{\odot}}\right)^{2} \\ \times \left(\frac{f_{ang}}{\eta} \frac{F_{bol}}{10^{-9} \mathrm{erg \ cm^{-2} \ s^{-1}}}\right)^{1/2} \frac{D}{3.5 \mathrm{kpc}} \ \mathrm{G\ cm}^{3}
\end{aligned}
\end{equation}
If we make the same assumptions about geometry and accretion efficiency (i.e. $k_{A}=1$, $f_{ang}=1$, and $\eta=0.1$), then $\mu\simeq1.62\times10^{26}$ G cm$^{3}$. This corresponds to a magnetic field strength of $B\simeq3.2\times10^{8}$ G at the magnetic poles for a NS of 10 km. Moreover, if we assume a different conversion factor $k_{A}=0.5$ \citep{long05} then the magnetic field strength at the poles would be $B\simeq1.0\times10^{9}$ G.
We note, however, that coherent pulsations have not been detected in 1RXS J1804, and the source is not identified as a pulsar.

However, if the disk does extend closer to the ISCO, then we can place a lower limit on the gravitational redshift from the NS surface, $z_{NS}$. Gravitational redshift is given by $z$+$1=1/\sqrt{1-2GM/R_{in}c^2}$. For $R_{in}=1.1$ ISCO, assuming a 1.4 M$_{\odot}$ NS and $a_{*}=0$, the $z_{NS}\geq0.2$ given that the neutron star is smaller than the radius of the disk. Our measurement for $R_{in}$ does extend down to 1 ISCO. If this were the case, then the $z_{NS}\geq0.22$ for a 12 km NS. 

To be thorough,  we fix the spin to 0.12 and 0.24 to test our assumption that the position of the inner disk radius changes slowly for low spin parameters as per \citealt{miller13}. The inner radius changes only marginally for the highest spin of $a_{*}=0.24$, confirming that our fits are not highly dependent upon the choice of spin parameter. 

\begin{figure}
\centering
\includegraphics[angle=270,width=8.4cm]{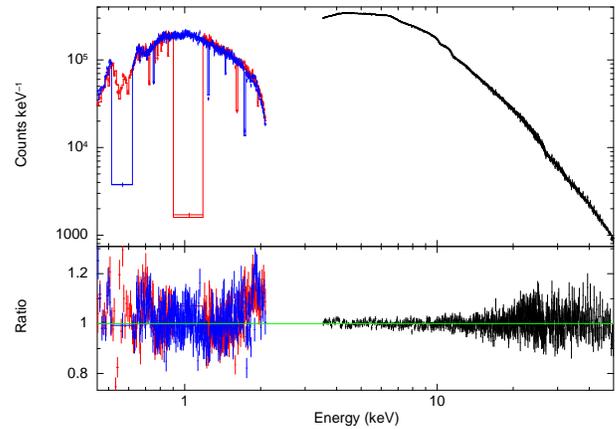}
\caption{Joint fit of RGS and $\emph{NuSTAR}$ with {\sc tbnew$*($relxill$+$gauss$+$gauss$+$relline)} ($\chi_{red}^{2}=1.34$). The additional {\sc relline} parameter is needed to model the He-like O VII line. Gaussians were fit at 10.1 keV and 11.5 keV to account for instrumental response features.
 Spectrum has been rebinned for plotting purposes. Parameter values can be seen in the fifth column of Table 1.}
\label{fig:joint}
\end{figure}  

\begin{table}
\caption{Double Relxill Fitting of $\emph{NuSTAR}$ and RGS}
\label{tab:relxill} 
\begin{center}
\begin{tabular}{llccc}
\hline
Component & Parameter & Values \\
\hline
{\sc tbnew}
&$N_\mathit{H} (10^{22})$
&$0.340\pm0.003$
\\
& $A_{O}$
&$1.689\pm0.03$
\\
{\sc relxill}
&$q_{in}=q_{out}$
&$<1.2$
\\
&$a_{*} '$
&0
\\
&$\mathit{i} (^{\circ})$
&$18.11_{-0.2}^{+0.3}$
\\
&$R_\mathit{in} (ISCO) $
&$1.13_{-0.13}^{+2.98}$
\\
&$R_\mathit{out} (R_\mathit{g}) '$ 
&400
\\
&$\mathit{z} '$
&0
\\
&$\Gamma$
&$<1.41$
\\
&$log(\xi)_{1}$
&$2.9\pm0.1$
\\
&$log(\xi)_{2}$
&$1.5\pm0.2$
\\
&$A_\mathit{Fe}$
&$<0.54$
\\
&$E_\mathit{cut} (keV)$
&$45.1\pm0.$
\\
&$\mathit{f}_\mathit{refl,1}$
&$0.2\pm0.1$
\\
&$\mathit{f}_\mathit{refl,2}$
&$0.22\pm0.02$
\\
&angleon$'$
&1
\\
&norm$_{1}$ $(10^{-2})$
&$4.73_{-1.3}^{+2.2}$
\\
&norm$_{2}$ $(10^{-2})$
&$4.68_{-1.0}^{+4.7}$
\\
\hline
&$\chi_\nu^{2}$(dof)
&1.36 (3962)
\\
\hline
$'$ = fixed
\end{tabular}

\medskip
Note.--- Errors are quoted at $\geq$ 90 \% confidence level. Setting angleon=1 takes the inclination into account when modeling reflection. A constant was allowed to float between $\emph{NuSTAR}$ and the RGS data. The $\emph{NuSTAR}$ was frozen at the value of 1.0, RGS1 was fit at $1.25\pm0.01$, and RGS2 was fit at $1.19\pm0.01$. 
\end{center}
\end{table}

\begin{figure}
\centering
\includegraphics[width=8.4cm]{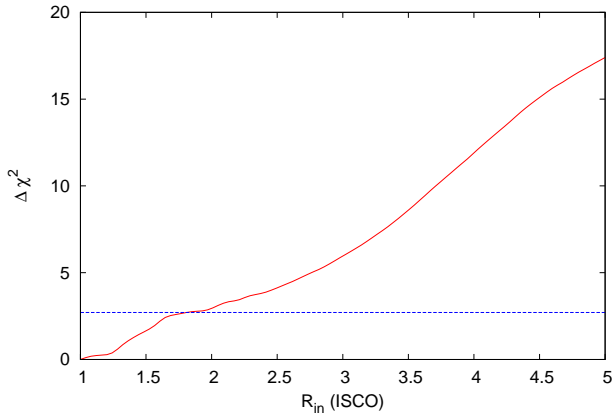}
\caption{ The change in goodness-of-fit versus inner disk radius for the $\emph{NuSTAR}+$RGS column in Table 1 taken over 50 evenly spaced steps generated with XSPEC \lq \lq steppar". The inner disk radius was held constant at each step while the other parameters were free to adjust. The blue dashed line shows the 90\% confidence level.}
\label{fig:joint}
\end{figure}  

\section{Discussion}
We have analyzed observations taken of the NS LMXB 1RXS J180408.9-34205 in the hard spectral state. We find clear evidence of a prominent iron line at 6.4 keV in the $\emph{NuSTAR}$ spectra and three reflection lines in the $\emph{XMM-Newton}$/RGS corresponding to N VII, O VII, and O VIII. This is the first time that a broad iron line has been detected in conjunction with multiple lower energy lines in the RGS.

We have found that when tying the $\emph{NuSTAR}$ and RGS data together, 1RXS J1804 has an $R_{in}\leq1.85$ ISCO with an inclination between $18^{\circ}$-$29^{\circ}$. This translates to $\leq22.2$ km for a 1.4 M$_{\odot}$ NS with $a_{*}=0$. Further, we are close to being able to map how ionization varies with radius in the disk. Individual fits to $\emph{NuSTAR}$ and RGS suggest that the O region may be located farther out in the disk than the Fe line region, but uncertainties do not exclude their mutual location. 

There are a growing number of NS LMXB with small inner radii measurements, such as Serpens X-1 (\citealt{miller13}; \citealt{chiang15}) and 4U 1608-52 \citep{degenaar15}. These objects were in a higher spectral state with softer photon indices that required the addition of thermal components in contrast to 1RXS J1804. Both objects had relativistic iron lines emerging from $R_{in}=6.2-7.2\ R_{g}$ \citep{miller13} for Serpens X-1 and $R_{in}=7-10\ R_{g}$ \citep{degenaar15} for 4U 1608-52.  Additionally, \citet{cackett10} looked at 7 other NS LMXBs  in addition to Serpens X-1. Nearly all of the objects were found to have an inner radius between $6-15\ R_{g}$. The relatively small inner radius measurement obtained for 1RXS J1804 suggests that the disk does not necessarily become highly truncated in the hard state. 
Similar behavior for the location of the inner radius was also seen for 4U 1705-44 \citep{disalvo15}. The disk radius in the hard state was comparable with the soft state. 

The ionization parameter for 1RXS J1804 is comparable to the values seen for other NS reflections studies for the iron line region, $2.3<log (\xi)<4.0$ (\citealt{cackett10}; \citealt{miller13}; \citealt{degenaar15}). However, other studies find lower ionization parameter values for a NS in the LHS (\citealt{Pap09}; \citealt{cackett09}; \citealt{dai10}; \citealt{disalvo15}). Though it is likely that some of the emergent flux in the iron line region for 1RXS J1804 is partially from a region of lower ionization. Again, we are close to tracing out the ionization as a function of disk radius and will push toward this goal in future work by achieving better overall data quality (higher signal to noise). We will focus on obtaining deeper observation with RGS for objects like 1RXS J1804 in the hard state with little to no thermal component to drown out the low energy reflection features. 

If the disk is not truncated at the stellar surface, it may be truncated at the magnetospheric radius. Using conservative methods, we find that the upper limit on $R_{in}$ implied by our fits would limit the magnetic field at the poles to $B\leq(0.3-1.0)\times10^{9}$ G (magnetic field strength at the poles is twice as large as at the magnetic equator). However, pulsations are not seen in this source and thus our estimate should be considered an upper limit.  An alternative explanation for disk truncation may be the presence of a boundary layer between the NS and inner disk (see \citealt{dai10} for more details).
\\
\\
\\
We thank the referee for the generous and helpful comments. We also thank Javier Garcia for insightful discussion regarding the {\sc relxill} model. EMC gratefully acknowledges support from the National Science Foundation through CAREER award number AST-1351222.
This research makes use of data from the $\emph{NuSTAR}$ mission, a project led by the California Institute of Technology, managed by the Jet Propulsion Laboratory, and funded by the National Aeronautics and Space Administration. We thank the $\emph{NuSTAR}$ Operations, Software and Calibration teams for support with the execution and analysis of these observations. This research has made use of the $\emph{NuSTAR}$ Data Analysis Software (NuSTARDAS) jointly developed by the ASI Science Data Center (ASDC, Italy) and the California Institute of Technology (USA). This research also made use of data obtained with $\emph{XMM-Newton}$, and ESA science mission with instruments and contributions directly funded by ESA Member States and NASA. LN wishes to acknowledge the Italian Space Agency (ASI) for financial support by ASI/INAF grant I/037/12/0-011/13.

\bibliographystyle{apj}
\bibliography{apj-jour,references}

\end{document}